\documentclass[12pt,british,nofootinbib,floatfix]{article}
\usepackage[T1]{fontenc}
\usepackage[latin9]{inputenc}
\usepackage{geometry}
\usepackage{amsmath}
\usepackage{amssymb}
\usepackage{graphicx}

\makeatletter

\providecommand{\tabularnewline}{\\}
\newcommand{\lyxdot}{.}

\numberwithin{equation}{section}
\numberwithin{figure}{section}
\numberwithin{table}{section}

\usepackage{cite}

\@ifundefined{showcaptionsetup}{}{%
 \PassOptionsToPackage{caption=false}{subfig}}
\usepackage{subfig}
\makeatother

\usepackage{babel}
\begin{document}

\title{Quasi-particle spectrum and entanglement generation after a quench
in the quantum Potts spin chain}

\author{O. Pomponio$^{1}$, L. Pristyák$^{1}$ and G. Takács$^{1,2}$\thanks{Corresponding author (email: takacsg@eik.bme.hu)}\\
{\small{}$^{1}$BME Department of Theoretical Physics }\\
{\small{}H-1111 Budapest, Budafoki út 8.}\\
{\small{}$^{2}$BME ``Momentum'' Statistical Field Theory Research
Group }\\
{\small{}H-1111 Budapest, Budafoki út 8.}}

\date{16th October 2018}
\maketitle
\begin{abstract}
Recently, a non-trivial relation between the quasi-particle spectrum
and entanglement entropy production was discovered in non-integrable
quenches in the paramagnetic Ising quantum spin chain. Here we study
the dynamics of analogous quenches in the quantum Potts spin chain.
Tuning the parameters of the system, we observe a sudden increase
in the entanglement production rate, which is shown to be related
to the appearance of new quasiparticle excitations in the post-quench
spectrum. Our results demonstrate the generality of the effect and
support its interpretation as the non-equilibrium version of the well-known
Gibbs paradox related to mixing entropy which appears in systems with
a non-trivial quasi-particle spectrum. 
\end{abstract}

\section{Introduction \label{sec:Introduction}}

A paradigmatic protocol for taking a quantum many-body system out
of equilibrium is provided by a quantum quench, which corresponds
to sudden change in the Hamiltonian. It is a protocol routinely engineered
in cold-atom experiments \cite{exp1,exp2,exp3,exp3a,exp4,exp5,exp6,exp7,exp8}
and provides a fruitful starting point to study non-equilibrium time
evolution of isolated quantum systems. When both the pre- and post-quench
Hamiltonians are translationally invariant, a quench starting from
an equilibrium (e.g. ground) state of the pre-quench system corresponds
to a situation with a uniform non-zero energy density under the post-quench
Hamiltonian, which is a highly excited configuration that can be considered
as a source of quasi-particle excitations \cite{calabrese-cardy}.
The subsequent time evolution can be considered as dynamics driven
by the quasi-particles created in the quench; the post-quench excitations
determine the spreading of correlation and entanglement in the system. 

Entanglement entropy is an important characteristics of the non-equilibrium
evolution and the stationary state resulting after a quench, and therefore
it has been studied extensively in recent years \cite{ent1,ent2,ent3,ent4,ent6,ent7,ent8,ent9,ent10,ent11,ent12,ent13,ent14,ent15}.
The growth of entanglement also has important implications for the
efficiency of computer simulations of the time evolution \cite{sim1,sim3,sim4,sim5}.
Recently it has become possible to measure entanglement entropy and
its temporal evolution in condensed matter systems \cite{exp2,Smeas1,Smeas2}.

For systems where interactions have a suitable fall-off with distance,
the quasi-particle propagation is limited by the existence of a maximum
speed $v_{\text{max}}$ called the Lieb-Robinson bound \cite{lieb-robinson}.
For the entanglement entropy $S_{\ell}$ of a subsystem of length
$\ell$ with the rest of system this results in an overall linear
growth of entanglement entropy $S_{\ell}(t)\sim t$ for times $t<\ell/2v_{\text{max}}$,
after which it becomes saturated as the subsystem approaches its stationary
state \cite{calcard_entropy}. The late time asymptotic value of entanglement
entropy of a large subsystem can also be interpreted as the usual
thermodynamic entropy \cite{exp2,calcard_entropy,integrable_entanglement_growth,deutsch,beugeling}.

In the regime dominated by the linear growth, entanglement generation
can be characterised by the mean entanglement entropy production rate
$\overline{\partial_{t}S}$, which naturally depends on the post-quench
spectrum and its quasi-particle content. For quenches to integrable
systems entanglement dynamics be computed from a recently developed
approach \cite{integrable_entanglement_growth,integrable_renyi,multi-particle_case,renyi_general_macrostates_integrable}.
The underlying quasi-particle description of entropy production describes
the initial state as a source of entangled quasi-particle pairs with
zero total momentum \cite{calabrese-cardy} the members of which propagate
to different parts of the system, resulting in the build-up of spatial
correlations and entanglement growth. This picture was explicitly
demonstrated for integrable quenches in the Ising spin chain \cite{calessfag}
and also forms the basis of a semi-classical approach for quantum
quenches \cite{semicl}, which is expected to be valid for sufficiently
small post-quench density even in the non-integrable case. It also
successfully describes entropy production in integrable systems \cite{integrable_entanglement_growth,multi-particle_case}
and leads to the following formula for the late time growth of the
entanglement entropy of a subsystem of size $\ell$ \cite{calcard_entropy,integrable_entanglement_growth,multi-particle_case}: 

\begin{equation}
S(t)\propto2t\sum_{n}\int_{2v_{n}t<\ell}dkv_{n}(k)f_{n}(k)+\ell\sum_{n}\int_{2v_{n}t>\ell}dkf_{n}(k)\;,\label{eq:semicl_entropy}
\end{equation}

\noindent where $n$ enumerates the different quasi-particle species,
$k$ is the momentum of the quasi-particles, $v_{n}(k)$ is their
velocity and $f_{n}(k)$ is a rate function describing the entropy
produced by quasi-particle pairs of species $n$ which depends on
their production rate. The restriction in the integrals reflects light-cone
propagation as a consequence of the Lieb\textendash Robinson bound.
For entanglement entropy between two halves of an infinite system
$\ell=\infty$, and so the second term describing saturation is absent,
while the integral in the first one has no restriction so Eq. (\ref{eq:semicl_entropy})
simplifies to 
\begin{equation}
S(t)\propto2t\sum_{n}\int dkv_{n}(k)f_{n}(k)\ .\label{eq:semicl_entropy_halfsystem}
\end{equation}
For free systems, the computation can be extended to a more general
class of initial states containing non-pair configurations \cite{no_pair}.
More recently, the generalised hydrodynamics approach to inhomogeneous
quenches \cite{ghd1,ghd2} was applied to compute entanglement in
free \cite{hydro_ent_free} and integrable \cite{hydro_ent_int} systems.

Much less is known about entanglement dynamics in quenches governed
by non-integrable post-quench dynamics. In the case of the quantum
Ising chain it was shown in recent studies that switching on an integrability
breaking longitudinal magnetic field $h_{x}$ leads to non-trivial
dynamical phenomena. In the ferromagnetic regime confinement suppresses
the usual linear growth of entanglement entropy and the corresponding
light-cone-like spreading of correlations after the quantum quench
\cite{Ising_confinement}. In contrast, there is no confinement in
the paramagnetic regime and thus entanglement entropy grows linearly
in time. Nevertheless the dependence of the mean entanglement entropy
production rate $\overline{\partial_{t}S}$ on the quench parameter
$h_{x}$ shows another kind of anomalous behaviour: a sudden increase
setting at the threshold value of $h_{x}$ where a new quasi-particle
excitation appears in the spectrum \cite{Ising_para}. Using the physical
interpretation of the asymptotic entanglement of a large subsystem
as the thermodynamic entropy of the stationary (equilibrium) state,
this can be recognised as arising from the contribution of mixing
entropy between the particle species, and so the effect can be interpreted
as a non-equilibrium manifestation of the Gibbs paradox.

The purpose of the present work is to demonstrate the same mechanism
in the quantum Potts spin chain, which is a generalisation of the
Ising case with three instead of two values for the spin variables.
Due to the higher symmetry of the chain, its spectrum and behaviour
is much richer, and a full exploration of the parameter space is out
of the scope of the present work. Instead we focus on the non-equilibrium
manifestation of the Gibbs mixing entropy analogous to the effect
found in \cite{Ising_para} to show that it generalises to the Potts
chain, and to support the interpretation advanced in \cite{Ising_para}.

The outline of the paper is as follows. Quenches in the paramagnetic
phase are considered first in Section \ref{sec:Quenches-in-the},
where it is shown that the effect observed in \cite{Ising_para} generalises
from the Ising to the Potts case. The determination and analysis of
spectrum in the paramagnetic phase are presented in Section \ref{sec:Spectrum-of-the}.
Section \ref{sec:Quasi-particle-spectrum-and} analyses the relation
between the time evolution and the quasi-particle spectrum, arguing
that the scenario proposed in \cite{Ising_para} holds for the Potts
case as well, and also discussing specific aspects where the Potts
model differs from the Ising case considered in \cite{Ising_para}.
Section \ref{sec:Conclusions} contains our conclusions.

\section{Quenches in the quantum potts spin chain \label{sec:Quenches-in-the}}

The 3-state Potts quantum spin chain is defined on the Hilbert space
\begin{equation}
\mathcal{H}=\bigotimes_{i=1}^{L}\left(\mathbb{C}^{3}\right)_{i}\label{eq:Hilbert_space}
\end{equation}
where $i$ labels the sites of the chain of length $L$. The quantum
space $\mathbb{C}^{3}$ at site $i$ has the basis $|\alpha\rangle$
with $\alpha=0,1,2$ corresponding to the spin degrees of freedom.
The dynamics is defined by the Hamiltonian 
\begin{equation}
H=-J\sum_{i=1}^{L}\left[\sum_{\alpha=0}^{2}\left(P_{i}^{\alpha}P_{i+1}^{\alpha}+h_{\alpha}P_{i}^{\alpha}\right)+g\tilde{P}_{i}\right]\label{eq:potts_chain_hamiltonian}
\end{equation}
where 
\begin{eqnarray}
P^{\alpha} & = & |\alpha\rangle\langle\alpha|-\frac{1}{3}\boldsymbol{1}_{3\times3}\label{eq:Palpha_and_Ptilde}\\
\tilde{P} & = & \frac{1}{3}\sum_{\alpha,\alpha'=0}^{2}\left(1-\delta_{\alpha\alpha'}\right)|\alpha\rangle\langle\alpha'|\nonumber 
\end{eqnarray}
and we assume periodic boundary conditions
\begin{equation}
P_{L+1}^{\alpha}\equiv P_{1}^{\alpha}\quad,\quad\tilde{P}_{L+1}\equiv\tilde{P}_{1}\label{eq:PBC}
\end{equation}
The parameters $h_{\alpha}$ and $g$ are dimensionless, while energy
(and by implication, time) units are specified by $J$. In all of
our subsequent numerical calculations we use units with $J=1$ and
also $\hbar=1$.

In the absence of the ``longitudinal'' magnetic fields $h_{\alpha}$,
the chain is invariant under the $\mathbb{S}_{3}$ permutation symmetry
of the three spin states $\alpha=0,1,2$ and it has a critical point
at $g=1$ corresponding to a phase transition between a paramagnetic
(PM) $g>1$ and ferromagnetic (FM) $g<1$ case. In the PM phase, there
is a single $\mathbb{S}_{3}$ invariant vacuum, while in the FM phase
there are three vacua that become degenerate in the infinite length
limit. The order parameter for the transition is given by the magnetizations
$m^{(\alpha)}=\langle P_{i}^{\alpha}\rangle$ and the quantum critical
point separating the phases can be described with a conformal field
theory (CFT) with central charge $c=4/5$.

\subsection{The quench protocol and the simulation procedure}

The non-equilibrium time evolution we study is defined by the following
quench protocol. The initial state is the ground state $|\Psi(0)\rangle$
of the pre-quench Hamiltonian
\begin{equation}
H_{\text{pre}}=-J\sum_{i=1}^{L}\left[\sum_{\alpha=0}^{2}\left(P_{i}^{\alpha}P_{i+1}^{\alpha}\right)+g\tilde{P}_{i}\right]\label{eq:initial_potts_hamiltonian}
\end{equation}
which is unique in the paramagnetic phase $g>1.$ We consider four
values $g=1.25,\;1.5,\;1.75$ and $2.0$, and the time evolution is
given by
\[
|\Psi(t)\rangle=e^{-iHt}|\Psi(0)\rangle
\]
where the post-quench Hamiltonian is given by (\ref{eq:potts_chain_hamiltonian_broken}):
\begin{equation}
H=-J\sum_{i=1}^{L}\left(\sum_{\alpha=0}^{2}P_{i}^{\alpha}P_{i+1}^{\alpha}+hP_{i}^{0}+g\tilde{P}_{i}\right)\label{eq:potts_chain_hamiltonian_broken-1}
\end{equation}
and we consider the time evolution as a function of $h$ which is
taken to be non-negative.

The time evolution is computed using the infinite volume time evolving
block decimation (iTEBD) algorithm \cite{vidal1}. Using translational
invariance, the many-body state is represented as the Matrix Product
State (MPS)

\[
|\Psi\rangle=\sum_{\ldots,s_{j},s_{j+1},\ldots}\cdots\Lambda_{o}\Gamma_{o}^{s_{j}}\Lambda_{e}\Gamma_{e}^{s_{j+1}}\cdots|\ldots,s_{j},s_{j+1},\ldots\rangle\;,
\]

\noindent where $s_{j}$ spans the local $3$-dimensional spin Hilbert
space, $\Gamma_{o/e}^{s}$ are $\chi\times\chi$ matrices associated
with the odd/even lattice site; $\Lambda_{o/e}$ are diagonal $\chi\times\chi$
matrices with the singular values corresponding to the bipartition
of the system at the odd/even bond as their entries. The many-body
state is initialised to the product state $|\Psi_{0}\rangle=\bigotimes\frac{1}{\sqrt{3}}(|0\rangle+|1\rangle+|2\rangle)$.
The ground state $|\Psi(0)\rangle$ was obtained by time-evolving
the initial state $|\Psi_{0}\rangle$ in imaginary time by the pre-quench
Hamiltonian (\ref{eq:initial_potts_hamiltonian}), using a second-order
Suzuki-Trotter decomposition of the evolution operator with imaginary
time Trotter step $\tau=10^{-3}$. Due to the presence of an energy
gap, an auxiliary dimension $\chi_{0}=81$ was sufficient to have
a very accurate result for the ground state.

The post-quench time evolution was obtained by evolving $|\Psi(0)\rangle$
with the post-quench Hamiltonian (\ref{eq:potts_chain_hamiltonian_broken-1})
in real time, again using a second-order Suzuki-Trotter decomposition
of the evolution operator with real time Trotter step $\delta t=0.005$.
To keep the truncation error small the auxiliary dimension was allowed
to grow up to $\chi_{\text{max}}=243$ which was sufficient to reach
a maximum time $T=40$. 

\subsection{Entanglement growth rate }

\begin{figure}[t]
\begin{centering}
\includegraphics{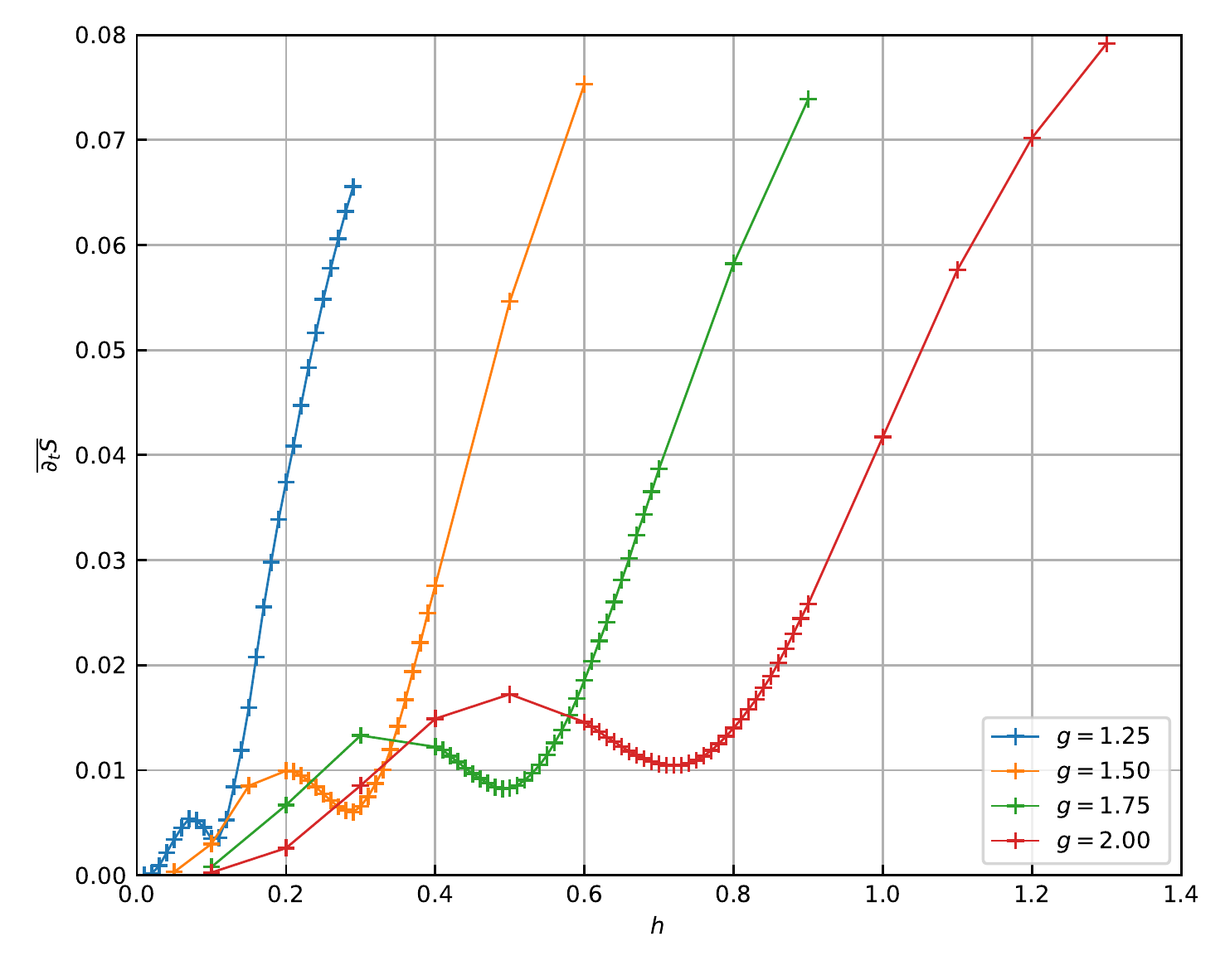}
\par\end{centering}
\caption{\label{fig:The-mean-entanglement} The mean entanglement entropy production
rate $\overline{\partial_{t}S}$ as a function of $h$ for $g=1.25,\;1.5,\;1.75$
and $2.0$.}
\end{figure}

The central issue of this work concerns the evolution of the half-system
entanglement entropy $S(t).$ This is defined by cutting the system
into two halves $\mathcal{H}$ and $\bar{\mathcal{H}}$ and introducing
the reduced density matrix
\[
\rho_{\mathcal{H}}(t)=\text{Tr}_{\bar{\mathcal{H}}}|\Psi(t)\rangle\langle\Psi(t)|
\]
Then the half-system entanglement entropy is given by 
\[
S(t)=-\text{Tr}_{\mathcal{H}}\rho_{\mathcal{H}}(t)\log\rho_{\mathcal{H}}(t)\,.
\]
\begin{figure}
\begin{centering}
\includegraphics{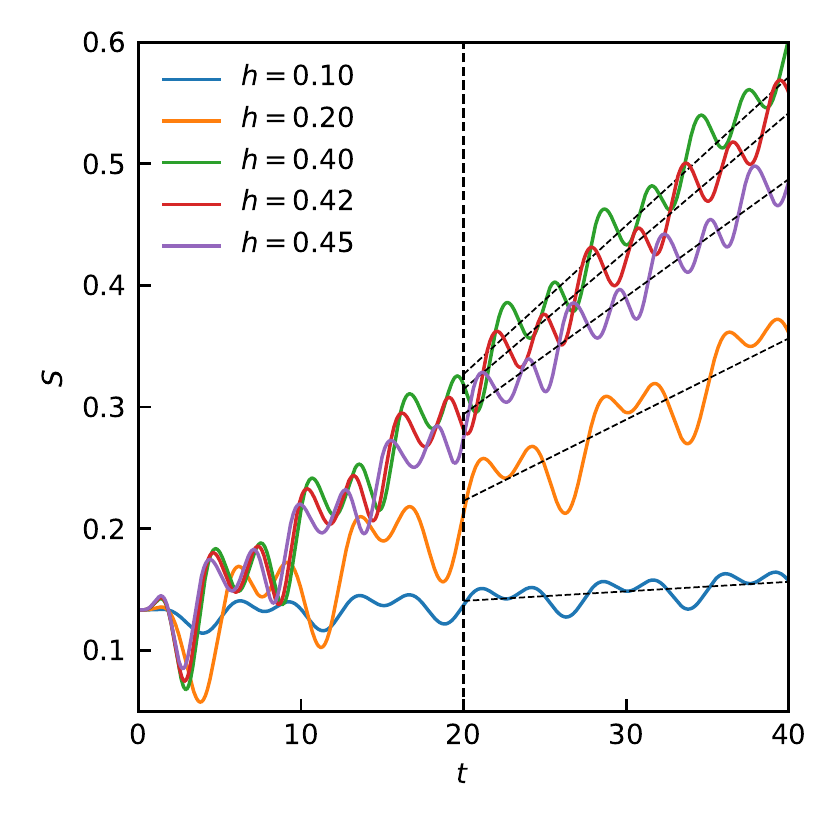}
\par\end{centering}
\caption{\label{fig:The-time-evolution}The time evolution of $S(t)$ for $g=1.75$
and $h=0.10$, $0.20$, $0.40$, $0.42$ and $0.45$, where time is
measured in units of $1/J$. The vertical line drawn at time $t$
shows the limit above which the average slope was extracted, the corresponding
fits are shown by the black dotted lines.}
\end{figure}
As illustrated in Fig. \ref{fig:The-time-evolution} after a relatively
short transient $S(t)$ shows a linear trend (with some slowly decaying
oscillations) as expected after a global quantum quench. A numerical
estimation of the mean entanglement entropy production rate $\overline{\partial_{t}S}$
was obtained by a linear fit of the iTEBD data in the time window
$20\leq t\leq40$. The dependence of $\overline{\partial_{t}S}$ on
$h$ for the values of the transverse field $g=1.25,\;1.5,\;1.75$
and $2.0$ is shown in Fig. \ref{fig:The-mean-entanglement}. It can
be seen clearly that $\overline{\partial_{t}S}$ has a local minimum
at a value $h_{\text{min}}$, the values of which are summarised in
the following table:
\begin{center}
\begin{tabular}{|c|c|c|c|c|}
\hline 
$g$ & $1.25$ & $1.5$ & $1.75$ & $2.0$\tabularnewline
\hline 
\hline 
$h_{\text{min}}$ & $0.10$ & $0.28$ & $0.49$ & $0.72$\tabularnewline
\hline 
\end{tabular}
\par\end{center}

This very peculiar, non-monotonous behaviour of $\overline{\partial_{t}S}$
was previously seen for quenches in the paramagnetic Ising spin chain
\cite{Ising_para} where it was explained by the effect of the quasi-particle
spectrum on the entanglement entropy production. In the following
we investigate the detailed dynamics of the Potts model to see whether
it confirms the scenario proposed in \cite{Ising_para}, which posited
that the reversal of the decreasing trend in $\overline{\partial_{t}S}$
at $h_{\text{min}}$ is due to the appearance of a new quasi-particle
excitation in the spectrum, which enhances entropy production by increasing
the number of species available. 

\section{Spectrum of the paramagnetic Potts spin chain \label{sec:Spectrum-of-the}}

\subsection{The case $h_{\alpha}=0$}

In the ferromagnetic phase the quasi-particle spectrum of the chain
consists of kink excitations $K_{\alpha\beta}$ connecting the vacua
according to the adjacency condition 
\begin{equation}
\alpha-\beta=\pm1\;\bmod\;3\label{eq:adjacency}
\end{equation}
with an obvious action of the permutation symmetry. 

In the paramagnetic phase the quasi-particle spectrum consists of
doubly degenerate magnons. Choosing two generators $\mathcal{C}$
and $\mathcal{T}$ for the group $\mathbb{S}_{3}$ which satisfy the
relations
\begin{equation}
\mathcal{T}^{3}=1\quad,\quad\mathcal{C}^{2}=1\quad,\quad\mathcal{CTC}=\mathcal{T}^{-1}\label{S3_generators}
\end{equation}
one can introduce a basis in the magnonic space with one-particle
states at fixed momentum given by $|A(k)\rangle$and $|\bar{A}(k)\rangle$.
They form the two-dimensional irreducible representation of $\mathbb{S}_{3}$
defined by the relations:
\begin{align}
\mathcal{T}|A(k)\rangle & =e^{2\pi i/3}|A(k)\rangle\nonumber \\
\mathcal{T}|\bar{A}(k)\rangle & =e^{-2\pi i/3}|\bar{A}(k)\rangle\nonumber \\
\mathcal{C}|A(k)\rangle & =|\bar{A}(k)\rangle\label{eq:S3_action}
\end{align}
For more information regarding the quasi-particle spectrum of the
chain we refer the interested reader to the work \cite{rapp} and
references therein. 

\subsection{The case $h_{\alpha}\protect\neq0$}

Switching on one or more longitudinal magnetic fields $h_{\alpha}$
leads to an explicit breaking of the symmetry group $\mathbb{S}_{3}$.
In the ferromagnetic case this results in confinement which is well-studied
in the scaling limit \cite{delfino_grinza,lepori,mesons,baryons,lencses}.
However, in this work we are interested in the paramagnetic phase,
and consider switching on one of the fields $h_{0}=h\neq0$ and keeping
$h_{1,2}=0$. Therefore our Hamiltonian is 
\begin{equation}
H=-J\sum_{i=1}^{L}\left(\sum_{\alpha=0}^{2}P_{i}^{\alpha}P_{i+1}^{\alpha}+hP_{i}^{0}+g\tilde{P}_{i}\right)\label{eq:potts_chain_hamiltonian_broken}
\end{equation}
This partially breaks the symmetry $\mathbb{S}_{3}$, leaving only
a $\mathbb{Z}_{2}$ subgroup intact. We choose the generator $\mathcal{C}$
to correspond to the unbroken subgroup, which in this case is generated
by the transformation swapping the spin directions $1$ and $2$.
Then for $h=0$ one can introduce the quasi-particle basis corresponding
to the eigenstates of $\mathcal{C}$
\begin{equation}
|A_{\pm}(k)\rangle=\frac{1}{\sqrt{2}}\left(|A(k)\rangle\pm|\bar{A}(k)\rangle\right)\label{eq:even_and_odd_qp}
\end{equation}
For $h=0$ they are degenerate, but for a non-zero $h$ the degeneracy
is lifted. As shown below, similarly to the case of the Ising spin
chain \cite{Ising_para}, for any fixed $g>1$ there is some critical
value $h_{\text{crit}}$ above which two $A_{+}$ quasi-particles
form a $\mathcal{C}$-even bound state $B$ which can formally be
written as a two-particle state with imaginary relative momentum
\begin{equation}
|B(k)\rangle\propto|A_{+}(k/2+i\kappa/2)A_{+}(k/2-i\kappa/2)\rangle\,.\label{eq:composite_bound_state}
\end{equation}
It is likely that the spectrum shows a larger variation when considering
the whole range of parameters $h_{\alpha}$ and $g$ (in the Ising
case, there also exist another bound state for larger values of $h$:
cf. \cite{Ising_para} for the spin chain, and \cite{zam_ising_spectr}
for the scaling field theory). However, in this work we restrict ourselves
to the regions $0\leq h\leq h_{\text{crit}}$ and $h_{\text{crit}}\lesssim h$,
and leave a more complete exploration of the parameter space for the
future.

\subsection{Quasi-particle dispersion relations for $h>0$}

\begin{figure}[t]
\centering{}\subfloat[$A_{+}$]{\noindent \begin{centering}
\includegraphics{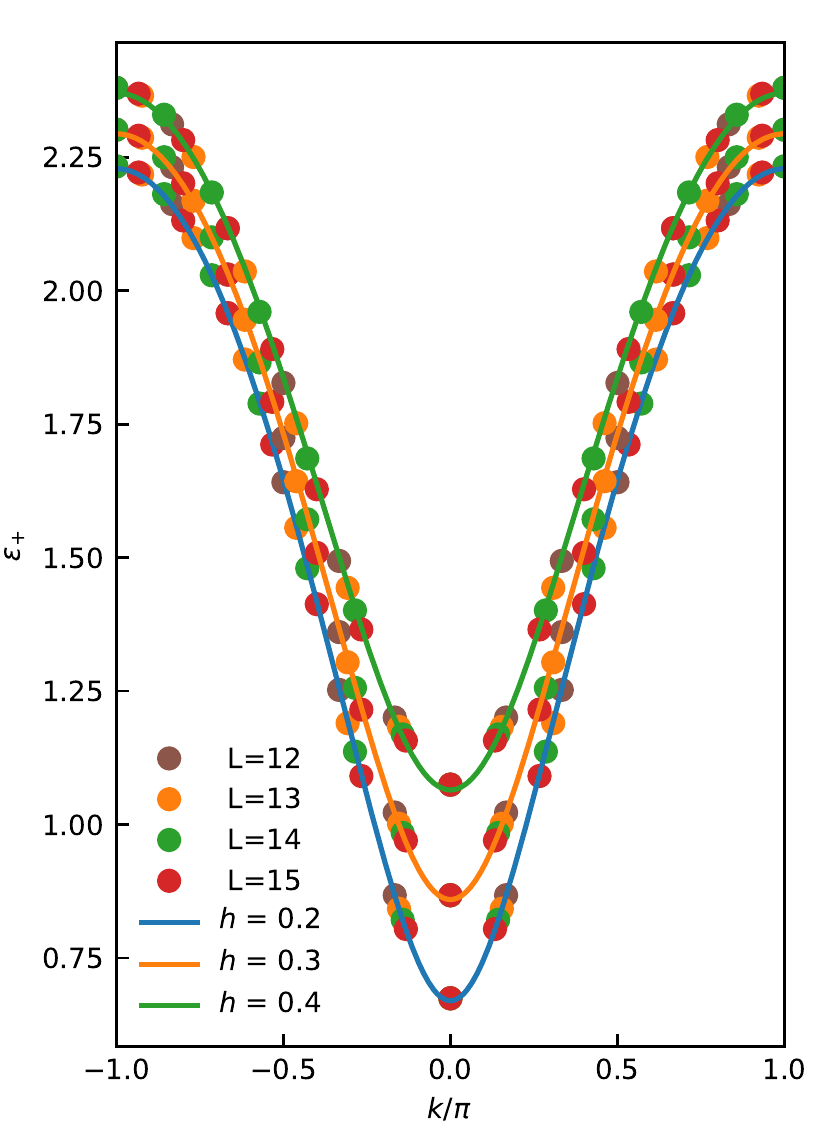}
\par\end{centering}
}\subfloat[$A_{-}$]{\noindent \begin{centering}
\includegraphics{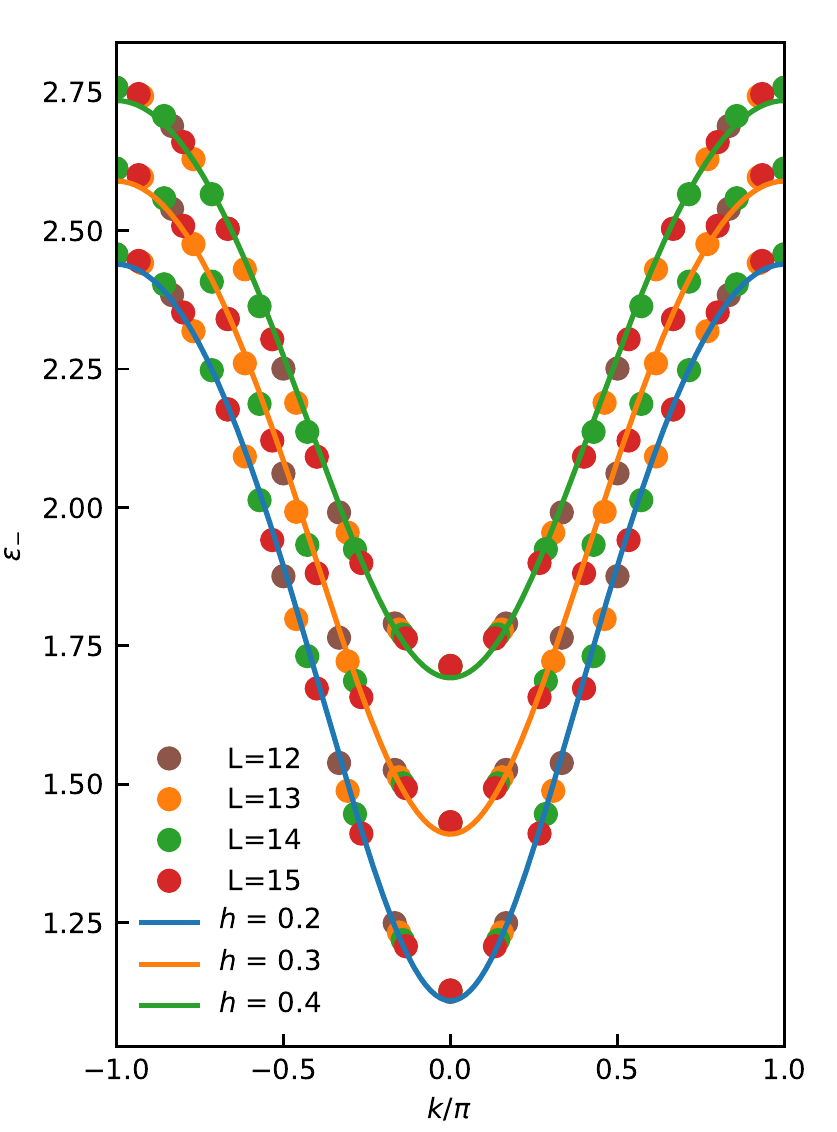}
\par\end{centering}
}\caption{\label{fig:Dispersion-relations-for}Dispersion relations for $A_{+}$
and $A_{-}$ for $g=1.5$ and $h=0.2$, $0.3$ and $0.4$. Energies
are shown in units of $J$, while momentum is shown in units of $1/a$,
where $a$ is the lattice spacing.}
\end{figure}

We determined the quasi-particle dispersion relations applying exact
diagonalisation of the Hamiltonian (\ref{eq:potts_chain_hamiltonian_broken}).
After determining a few hundred states at the bottom of the spectrum,
they were sorted into bins containing energy levels that are degenerate
within numerical precision. Most of the eigenvalues appear in degenerate
pairs of states with opposite total momentum $k$ and $-k$, with
the exception of singlets with momenta $k=0$ and $k=\pi$. Momenta
of states can be obtained by diagonalising the shift operator $\mathcal{S}$
mapping site $i$ to site $i+1\bmod L$ within the bins. In the paramagnetic
phase, the ground state is an isolated singlet, followed by two branches
of one particle states corresponding to momenta 
\begin{equation}
k_{n}=n\frac{2\pi}{L}\:,\:n=\left[-\frac{L}{2}\right]+1,\dots,\left[\frac{L}{2}\right]\label{eq:quantised_momenta}
\end{equation}
where the two branches are distinguished by the eigenvalue of $\mathcal{C}$
which corresponds to the unbroken $\mathbb{Z}_{2}$. We computed the
quasi-particle branches for chain lengths $L$ from $12$ to $15$.
An example result is shown in Fig. \ref{fig:Dispersion-relations-for}.

\begin{figure}[t]
\begin{centering}
\subfloat[$g=1.25$]{\begin{centering}
\includegraphics{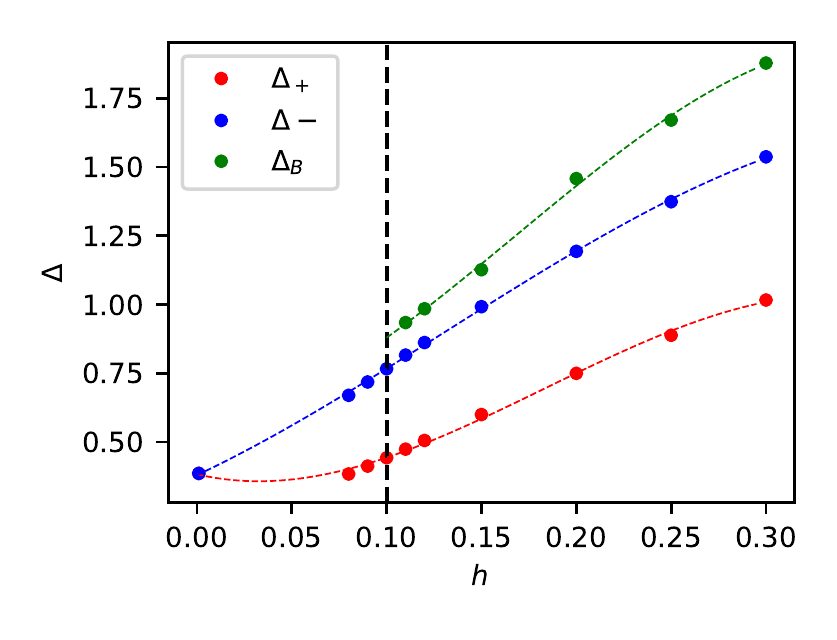}
\par\end{centering}
}\subfloat[$g=1.5$]{\begin{centering}
\includegraphics{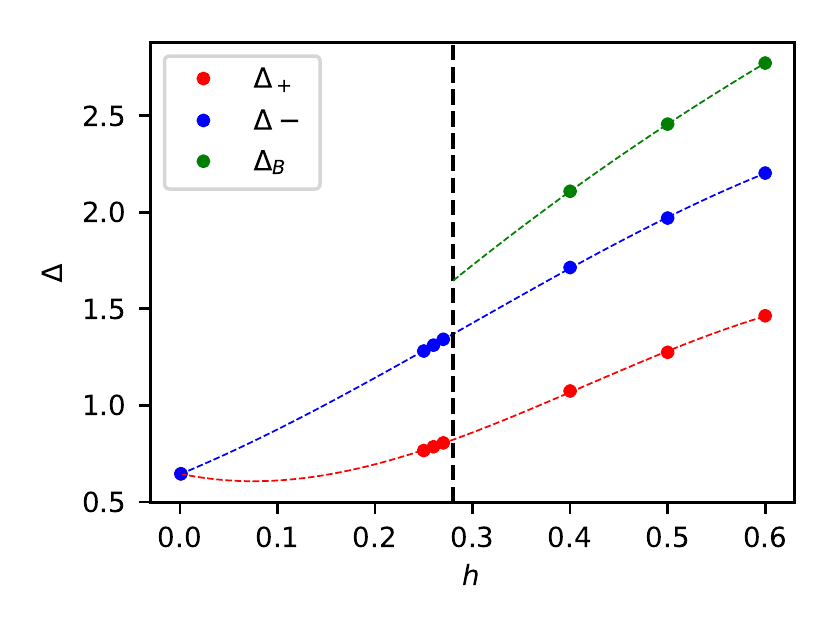}
\par\end{centering}
}
\par\end{centering}
\begin{centering}
\subfloat[$g=1.75$]{\begin{centering}
\includegraphics{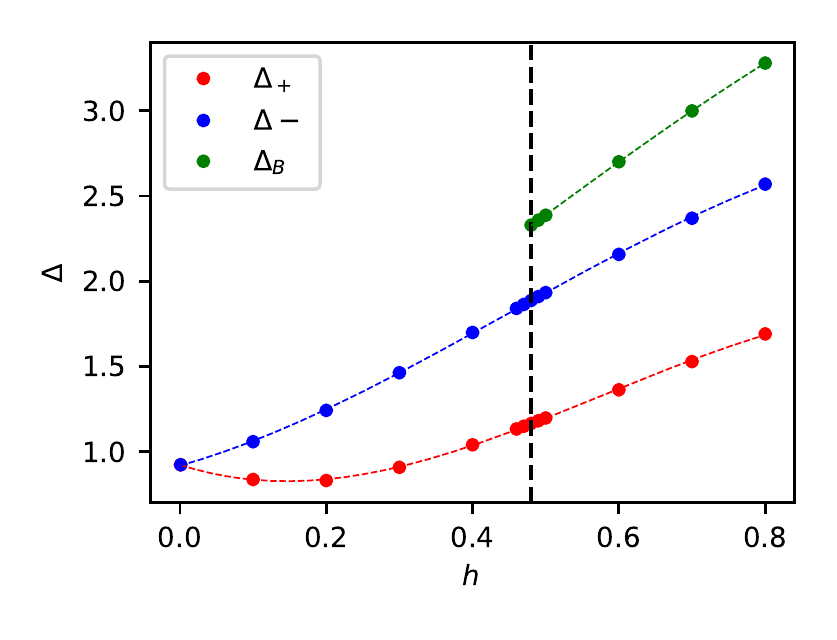}
\par\end{centering}
}\subfloat[$g=2$]{\begin{centering}
\includegraphics{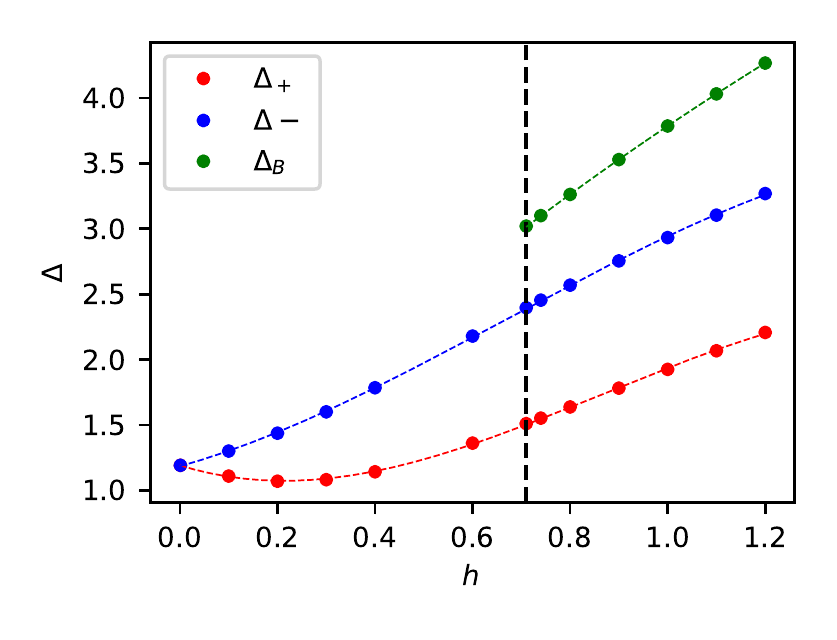}
\par\end{centering}
}
\par\end{centering}
\caption{\label{fig:Quasi-particle-gaps-}Quasi-particle gaps $\Delta_{\pm}$
and $\Delta_{B}$ (in units of $J$) as functions of $h$. The vertical
dashed line shows the threshold value of $h$ above which the bound
state quasi-particle $B$ exists.}
\end{figure}

\begin{figure}[t]
\begin{centering}
\subfloat[$g=1.25$]{\begin{centering}
\includegraphics{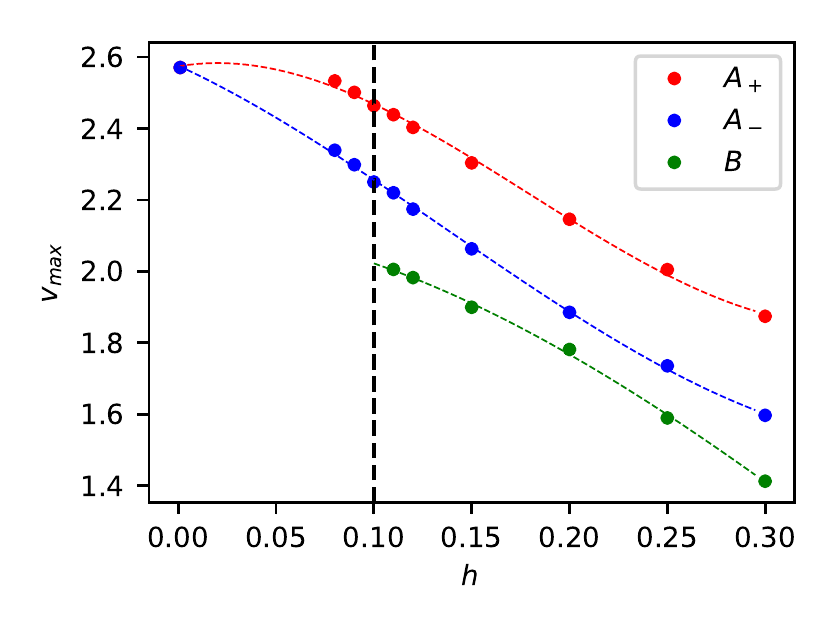}
\par\end{centering}
}\subfloat[$g=1.5$]{\begin{centering}
\includegraphics{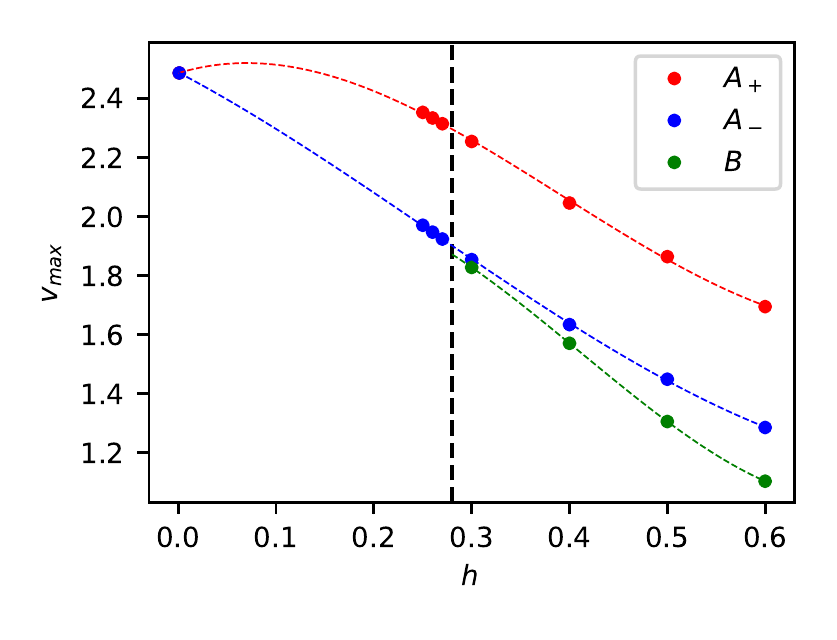}
\par\end{centering}
}
\par\end{centering}
\begin{centering}
\subfloat[$g=1.75$]{\begin{centering}
\includegraphics{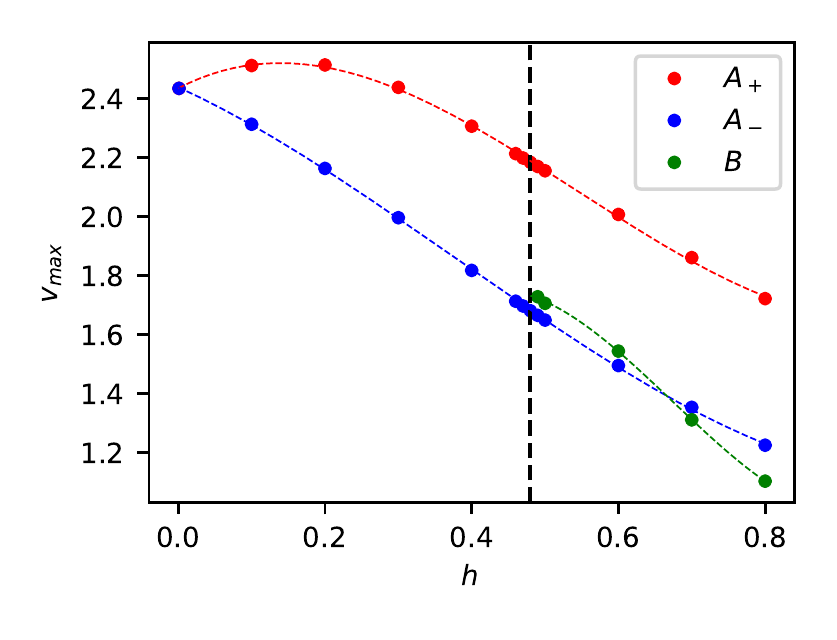}
\par\end{centering}
}\subfloat[$g=2$]{\begin{centering}
\includegraphics{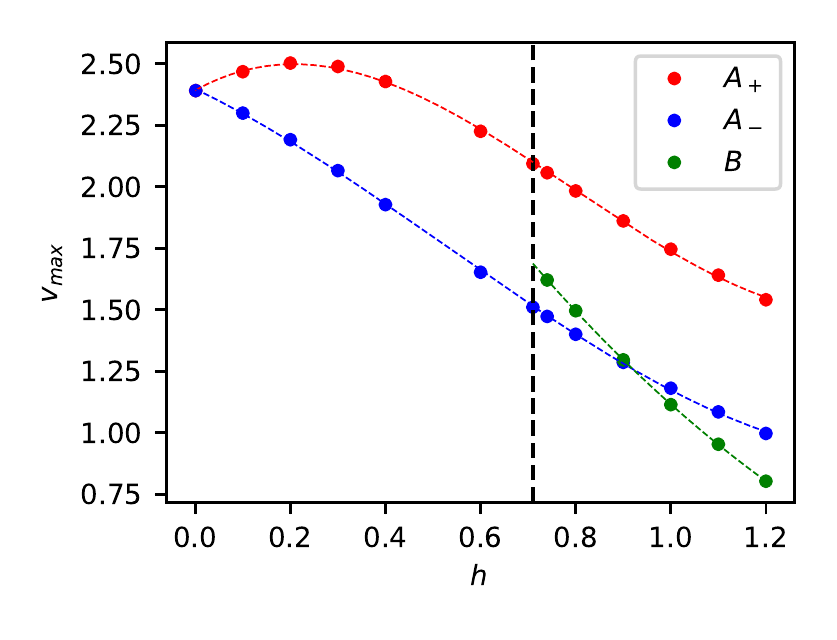}
\par\end{centering}
}
\par\end{centering}
\caption{\label{fig:Lieb-Robinson-velocities-}Lieb-Robinson velocities $v_{\mathrm{max}\pm}$
and $v_{\mathrm{max}B}$ as functions of $h$. The vertical dashed
line shows the threshold value of $h$ above which the bound state
quasi-particle $B$ exists. Velocities are shown in units of $Ja$,
where $a$ is the lattice spacing.}
\end{figure}

It turns out that dependence on the finite size $L$ is very weak,
so one can treat the results from different chain lengths $L$ as
sampling the same (infinite volume) dispersion relations $\epsilon_{\pm}(k)$.
In addition, the fitting functions 
\begin{equation}
\epsilon_{\pm}(k)=\sqrt{a_{\pm}+b_{\pm}\cos k}\label{eq:fitting_functions}
\end{equation}
inspired by the free fermion dispersion relation provide an excellent
description of the numerical data. The fits can be used to determine
both the gaps
\begin{equation}
\Delta_{\pm}=\sqrt{a_{\pm}+b_{\pm}}\label{eq:gaps_from_dispersion_relation}
\end{equation}
and the Lieb-Robinson (LR) velocities
\begin{equation}
v_{\mathrm{max}\pm}=\max_{k}\frac{\partial\epsilon_{\pm}}{\partial k}\label{eq:LR_velocity}
\end{equation}
To find the bound state threshold, we use a different approach for
the determination of the gap that leads to a much more precise result.
Note that for each chain length one can obtain the value of the gap
$\Delta_{\pm}(L)$ from the energy of the first/second zero momentum
excited state relative to the ground state. Using the theory of finite
size effects \cite{luscher_onept} one can then extrapolate these
to infinite volume using the fitting functions 
\begin{equation}
\Delta_{\pm}(L)=\Delta_{\pm}+\gamma_{\pm}e^{-\mu_{\pm}L}\label{eq:gap_volume_dependence}
\end{equation}
For the choices of the transverse field $g=1.25,\;1.5,\;1.75$ and
$2.0$, the dependence of the gaps and LR velocities on the longitudinal
field $h$ are shown in Figs. \ref{fig:Quasi-particle-gaps-} and
\ref{fig:Lieb-Robinson-velocities-}, respectively.

\subsection{Bound state threshold}

\subsubsection{Determination of $h_{\text{crit}}$}

For $h<h_{\text{crit}}$, the part of the spectrum above the two quasi-particle
branches consists of many-particle states called the continuum. Since
$\Delta_{+}<\Delta_{-}$ for $h>0$, the lowest lying levels are two-particle
states\footnote{In fact, this state can hybridize with $A_{-}A_{-}$ two-particle
states, but it does not change the subsequent considerations and so
we omit this term for simplicity.} 
\[
|A_{+}(k/2+q/2)A_{+}(k/2-q/2)\rangle
\]
of total momentum $k$ taking the values (\ref{eq:quantised_momenta}),
while $q$ is their relative momentum, which is quantised differently
due to interaction effects involving the scattering phase shift \cite{luscher_2particle}.
The lowest lying even state above $|A_{+}(0)\rangle$ corresponds
to a zero-momentum state
\[
|A_{+}(q_{\text{min}}/2)A_{+}(-q_{\text{min}}/2)\rangle
\]
where $q_{\text{min}}$ is the smallest allowed value for the relative
momentum $q$. When $h$ approaches $h_{\text{crit}}$, $q_{\text{min}}$
goes to $0$ and for $h>h_{\text{crit}}$ it turns imaginary according
to the standard quantum mechanical relation between scattering and
bound states, with the state becoming identical to 
\[
|B(0)\rangle
\]
i.e. a zero-momentum level with a single $B$ quasi-particle. Denoting
the energy gap of this level by $\Delta_{B}$ one has 
\begin{align}
\Delta_{B} & >2\Delta_{+}\quad h<h_{\text{crit}}\nonumber \\
\Delta_{B} & =2\Delta_{+}\quad h=h_{\text{crit}}\label{eq:hcrit_criteria}\\
\Delta_{B} & <2\Delta_{+}\quad h>h_{\text{crit}}\nonumber 
\end{align}
which makes possible the determination of $h_{\text{crit}}$. 

\begin{figure}[t]
\begin{centering}
\subfloat[$h=0.6$]{\begin{centering}
\includegraphics{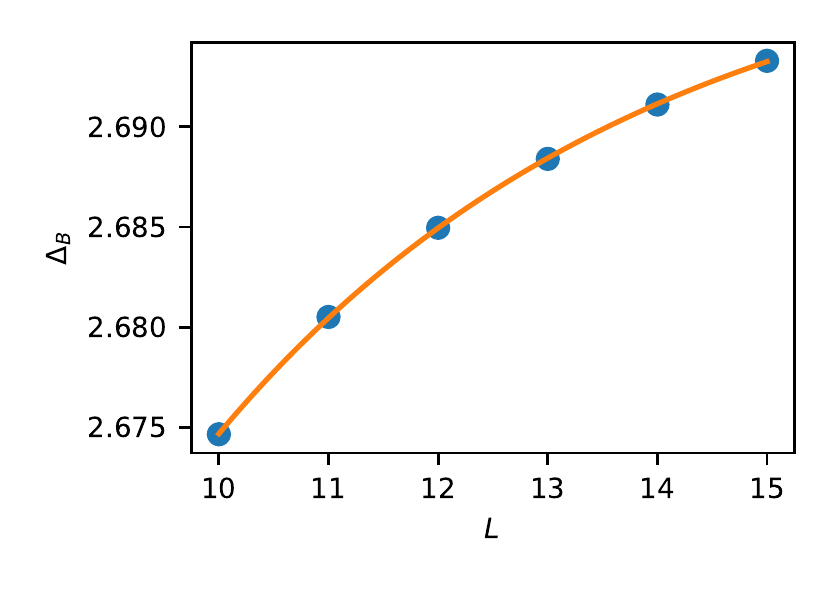}
\par\end{centering}
}\subfloat[$h=0.49$]{\begin{centering}
\includegraphics{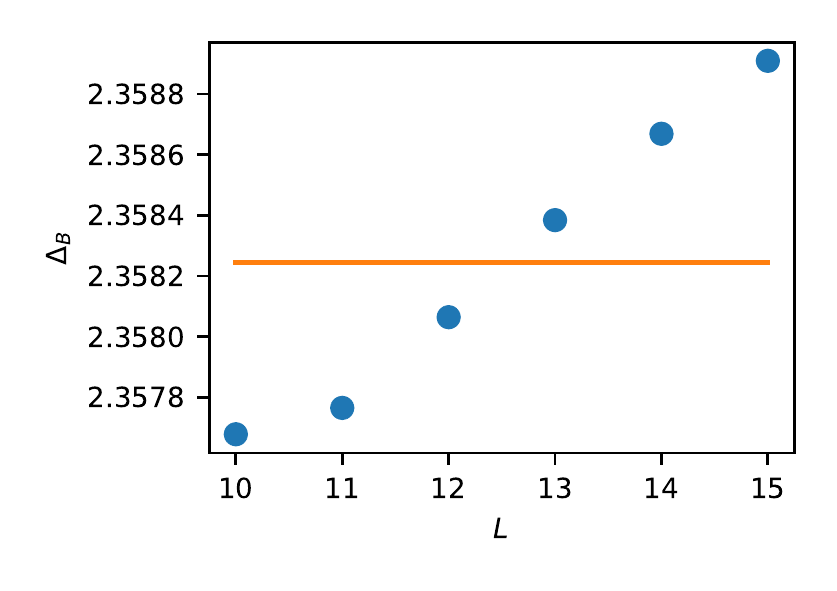}
\par\end{centering}
}
\par\end{centering}
\caption{\label{fig:Finite-size-extrapolation} Finite size extrapolation for
the gap $\Delta_{B}$ (in units of $J$). Note that well above the
threshold value of $h\approx0.49$ the exponential fit works very
well (a), while at the threshold it completely misses (b). However,
comparing (a) abd (b) shows that the range of variation of $\Delta_{B}$
is much smaller when at the threshold, corresponding to the vanishing
of the leading order finite size correction.}
\end{figure}

Finite size effects can be eliminated using the exponential extrapolation
according to the leading order finite size dependence predicted in
\cite{luscher_onept} 
\begin{equation}
\Delta_{B}(L)=\Delta_{B}+\gamma_{B}e^{-\mu_{B}L}\label{eq:Bgap_extrapol}
\end{equation}
when $h>h_{\text{crit}}$. Here $1/\mu_{B}$ is a length scale corresponding
to the spatial extension of the $A_{+}A_{+}$ bound state wave function,
which diverges at $h=h_{\text{crit}}$ and so the simple exponential
extrapolation prescribed by (\ref{eq:Bgap_extrapol}) becomes impossible
in the vicinity of the threshold $h_{\text{crit}}$, as illustrated
in Fig. \ref{fig:Finite-size-extrapolation}. However, in that case
$\gamma_{B}$ vanishes as well since it corresponds to the effective
coupling between two $A_{+}$ particles which changes sign from attractive
to repulsive and so vanishes at the threshold, so the finite volume
dependence is much weaker as it is determined by subleading corrections.

\begin{figure}[t]
\begin{centering}
\subfloat[$g=1.25$]{\begin{centering}
\includegraphics{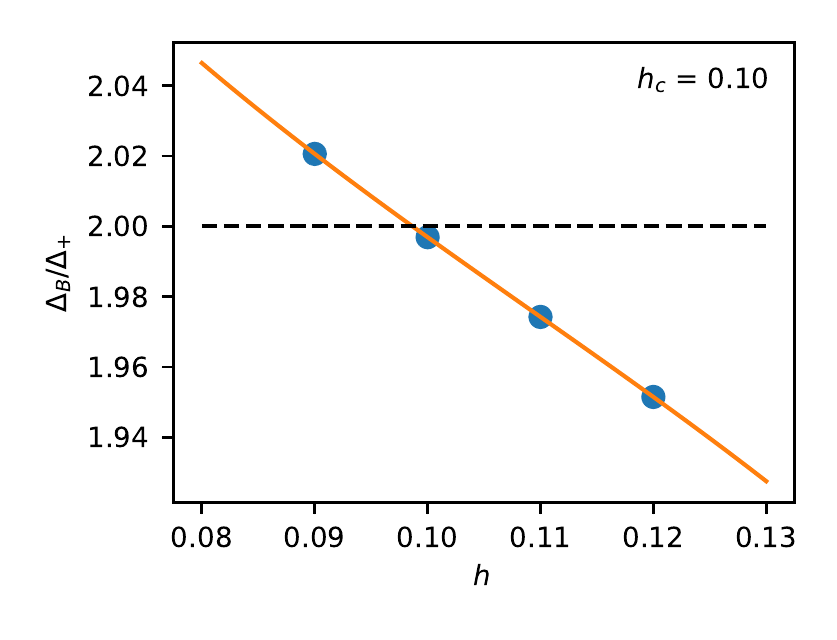}
\par\end{centering}
}\subfloat[$g=1.5$]{\begin{centering}
\includegraphics{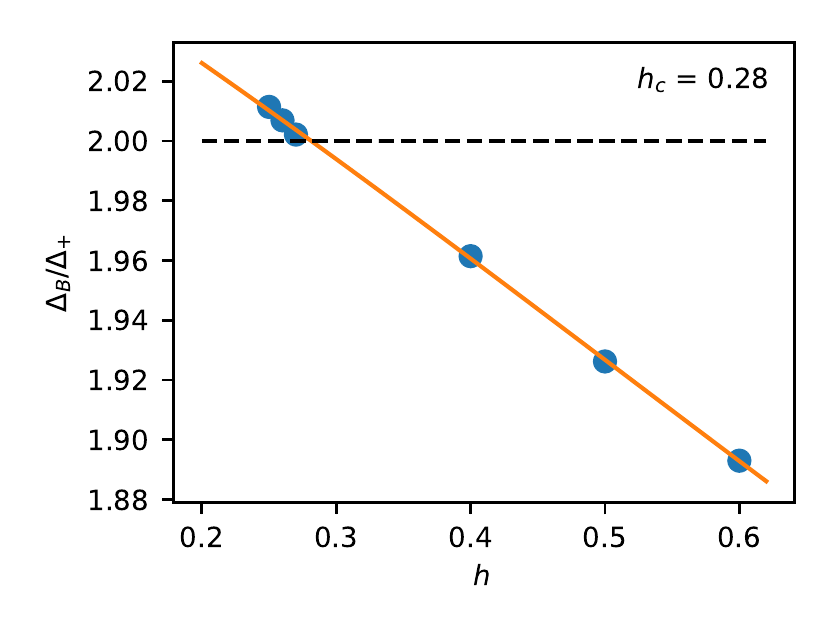}
\par\end{centering}
}
\par\end{centering}
\begin{centering}
\subfloat[$g=1.75$]{\begin{centering}
\includegraphics{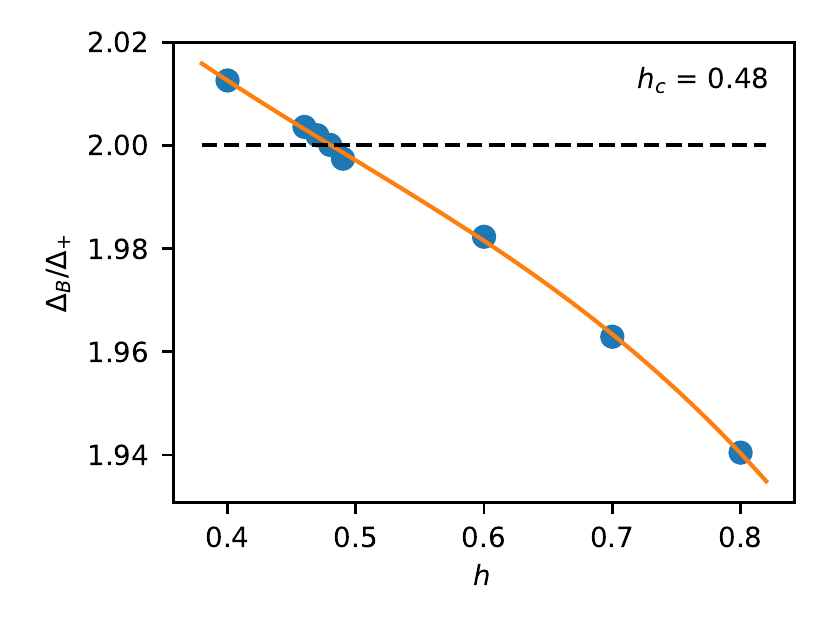}
\par\end{centering}
}\subfloat[$g=2$]{\begin{centering}
\includegraphics{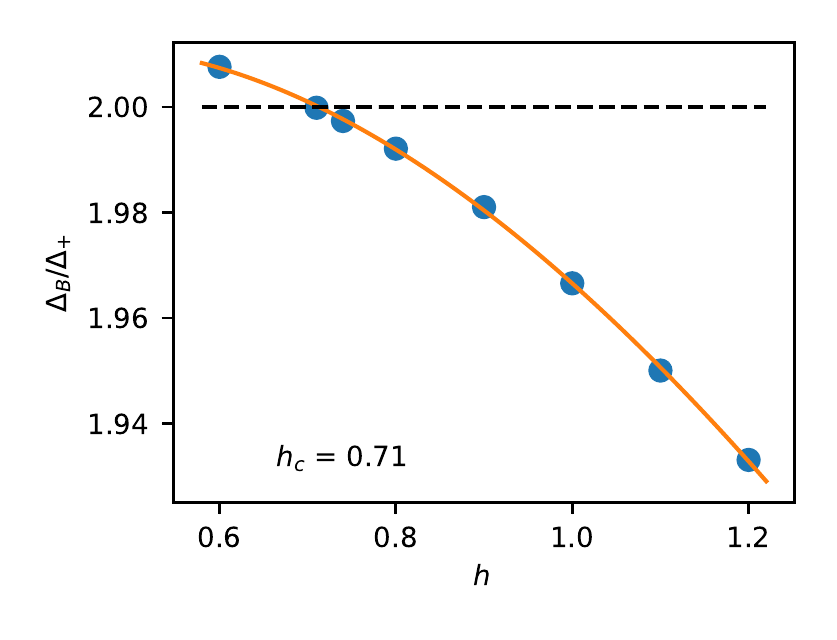}
\par\end{centering}
}
\par\end{centering}
\caption{\label{fig:Determining--from}Determining $h_{\text{crit}}$ from
$\Delta_{B}/\Delta_{+}$ as a function of $h$.}
\end{figure}

For $h<h_{\text{crit}}$ the energy level is a scattering state and
volume dependence is different (decaying as a power in $L$) and much
better numerical data are necessary in order to describe it in terms
of scattering characteristics \cite{luscher_2particle}. However,
since we are not interested in finding the actual value of the energy
level in that range we can simply fit it by the same function (\ref{eq:Bgap_extrapol})
to keep our procedure uniform.

The value of $h_{\text{crit}}$ can be determined by plotting $\Delta_{B}/\Delta_{+}$
as a function of $h$ and finding the value where it crosses $2$,
as shown in Fig. \ref{fig:Determining--from}.

\subsubsection{Dispersion relation for $B$ }

\begin{figure}[t]
\begin{centering}
\includegraphics{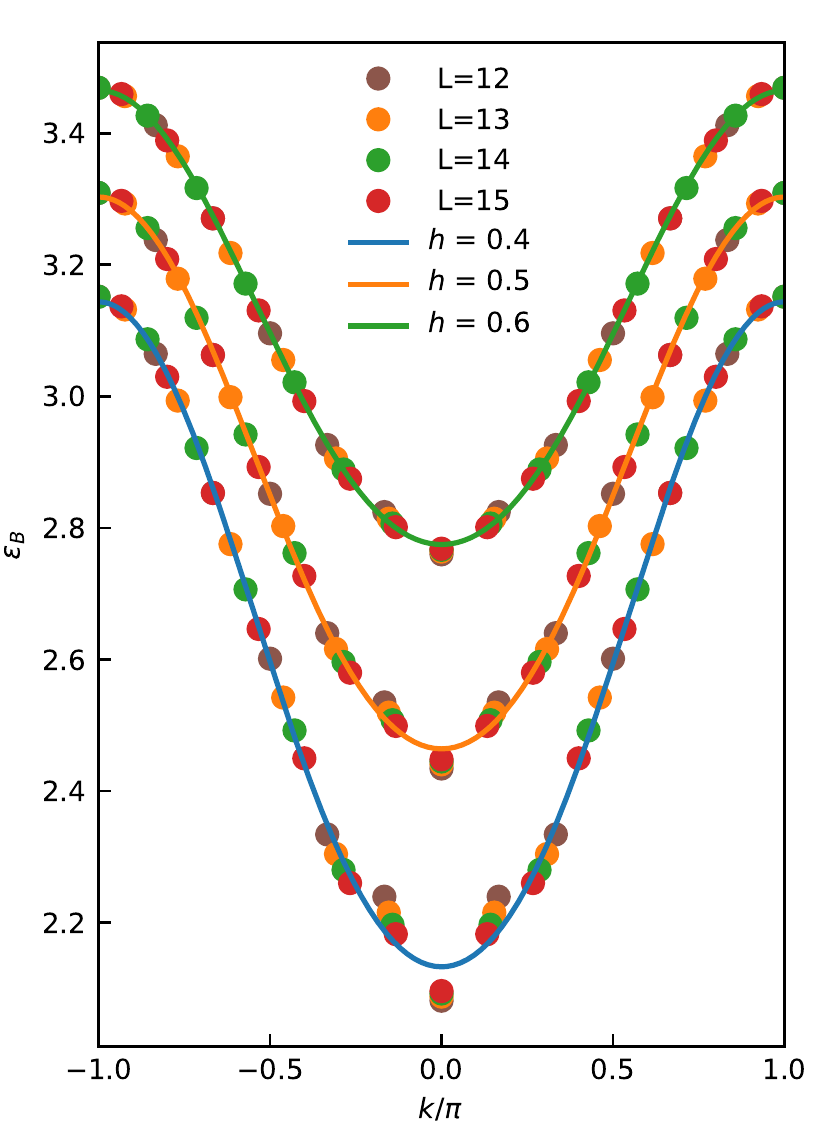}
\par\end{centering}
\caption{\label{fig:Dispersion-relation-for-B}Dispersion relation for $B$
at $g=1.5$ and $h=0.4$, $0.5$ and $0.6$. Energies are shown in
units of $J$, while momentum is shown in units of $1/a$, where $a$
is the lattice spacing.}
\end{figure}

In the regime $h>h_{\text{crit}}$ the states with momenta (\ref{eq:quantised_momenta})
lying just above the two quasi-particle branches $A_{\pm}$ correspond
to single-quasi-particle states $|B(k_{n})\rangle$, and their energies
allow the determination of the dispersion relation of $B$ as shown
in Fig. \ref{fig:Dispersion-relation-for-B}. Just as it was noted
in the case of the Ising chain \cite{Ising_para}, the dispersion
relation can be fitted well with the function
\[
\epsilon_{B}(k)=\sqrt{a_{B}+b_{B}\cos k+c_{B}\cos2k}
\]
from which it is possible to determine both the gap $\Delta_{B}=\sqrt{a_{B}+b_{B}+c_{B}}$
and the Lieb-Robinson velocity $v_{\mathrm{max}\pm}={\displaystyle \max_{k}}{\displaystyle \frac{\partial\epsilon_{\pm}}{\partial k}}$.
Just as in the case of the ``elementary'' quasi-particles $A_{\pm}$,
the gap $\Delta_{B}$ can also be determined using the extrapolation
(\ref{eq:Bgap_extrapol}) which leads to a more accurate result. It
is also clear from Fig. \ref{fig:Dispersion-relation-for-B} that
moving closer to the threshold i.e. for smaller $h$, when the quasi-particle
$B$ becomes more weakly bound the finite size dependence becomes
stronger, which is in fact expected from \ref{eq:Bgap_extrapol} since
when the spatial extension of the bound state wave function increases,
the exponent $\mu_{B}$ becomes smaller.

\section{Quasi-particle spectrum and non-equilibrium time evolution \label{sec:Quasi-particle-spectrum-and}}

\subsection{Time evolution of magnetisation}

\begin{figure*}
\begin{centering}
\subfloat[$m_{1}(t)$ and its FPS for $h=0.06$]{\begin{centering}
\includegraphics{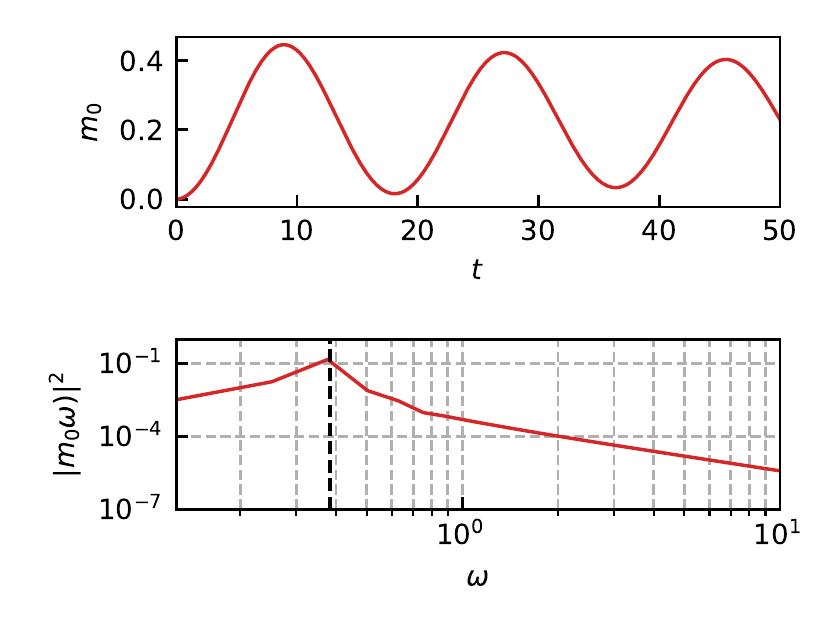}
\par\end{centering}
}\subfloat[$\tilde{m}(t)$ and its FPS for $h=0.06$]{\begin{centering}
\includegraphics{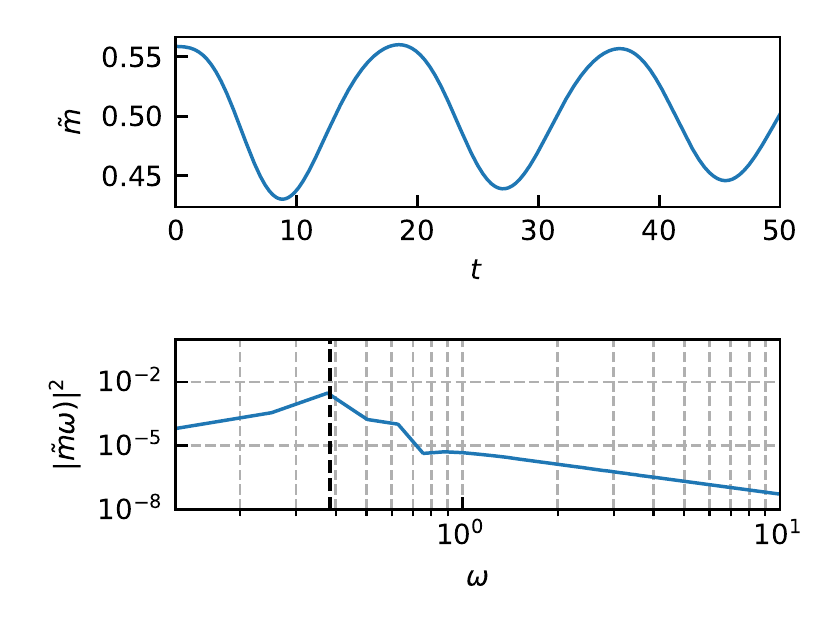}
\par\end{centering}
}
\par\end{centering}
\begin{centering}
\subfloat[$m_{1}(t)$ and its FPS for $h=0.1$]{\begin{centering}
\includegraphics{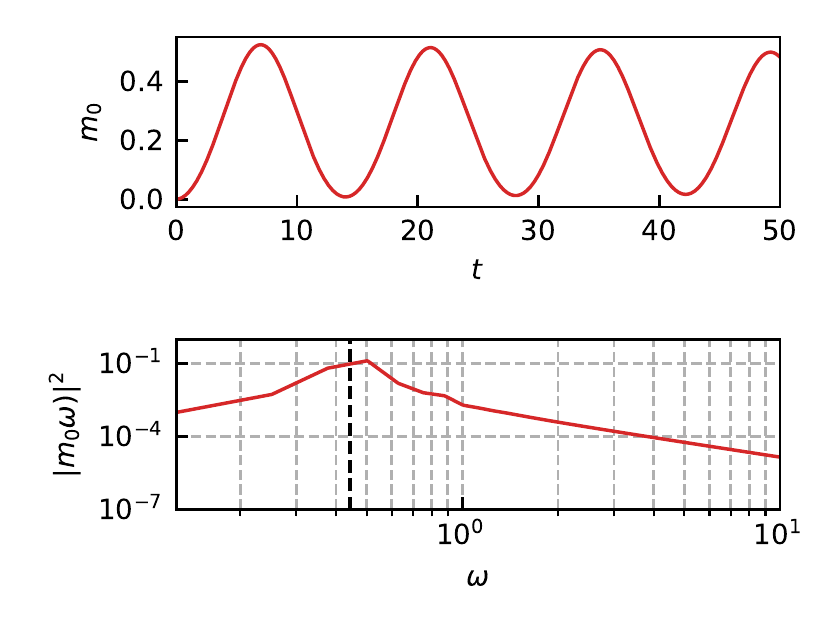}
\par\end{centering}
}\subfloat[$\tilde{m}(t)$ and its FPS for $h=0.1$]{\begin{centering}
\includegraphics{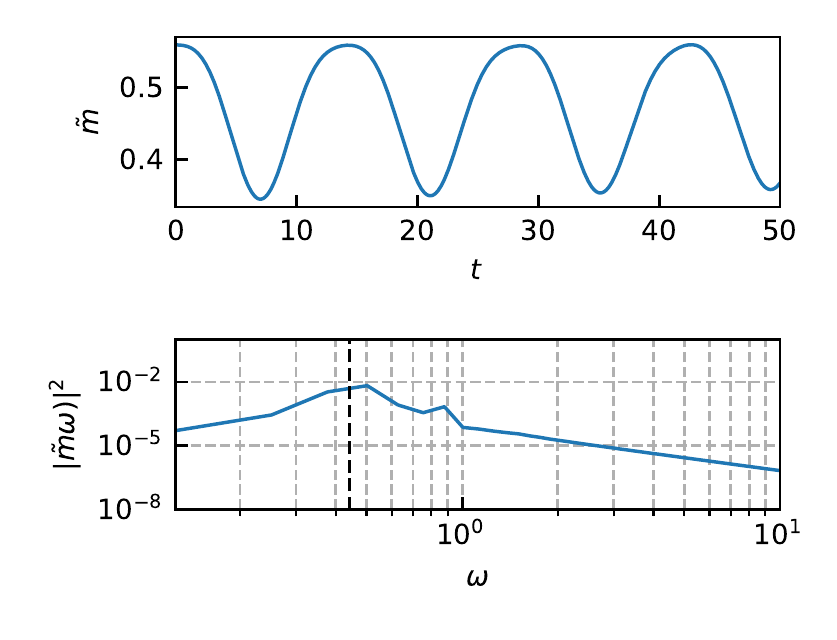}
\par\end{centering}
}
\par\end{centering}
\begin{centering}
\subfloat[$m_{1}(t)$ and its FPS for $h=0.12$]{\begin{centering}
\includegraphics{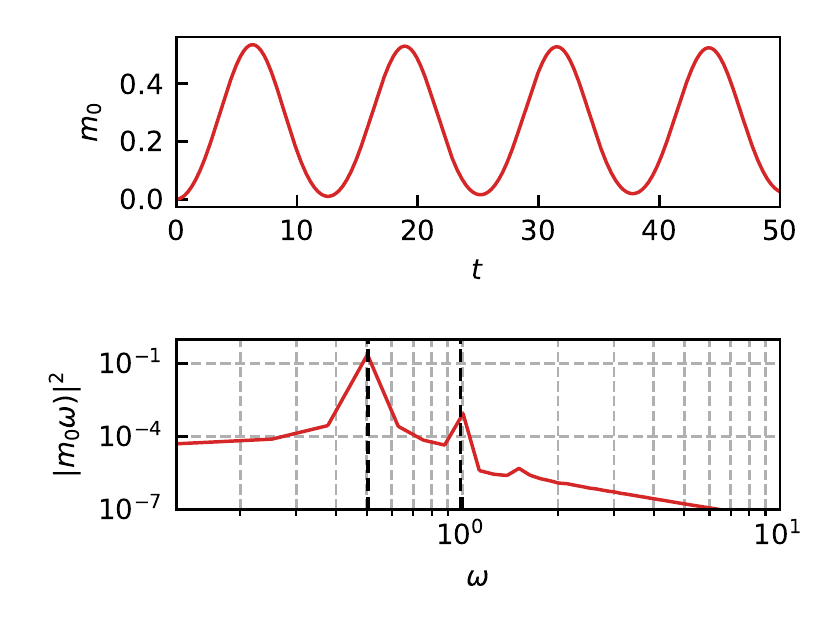}
\par\end{centering}
}\subfloat[$\tilde{m}(t)$ and its FPS for $h=0.12$]{\begin{centering}
\includegraphics{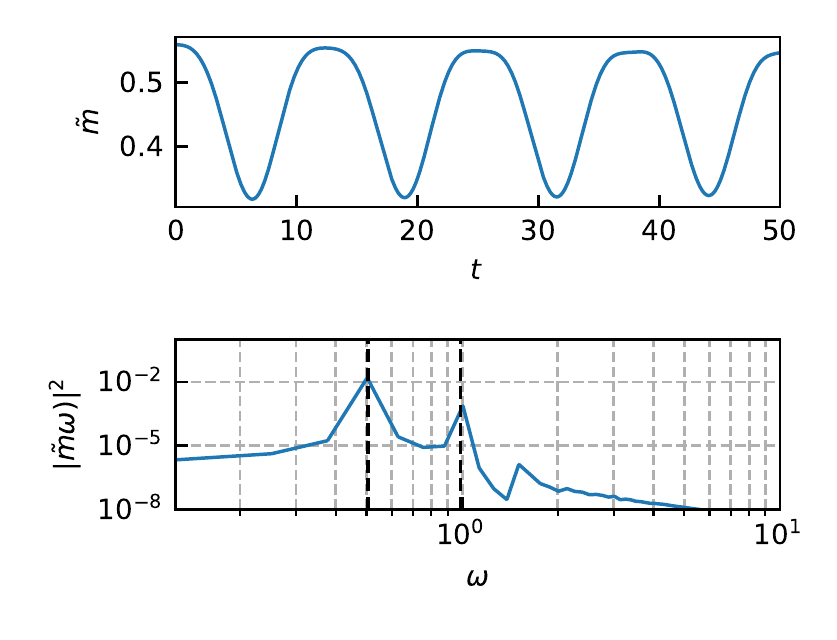}
\par\end{centering}
}
\par\end{centering}
\caption{\label{fig:Time-evolution-of}Time evolution of the longitudinal and
transverse magnetisations $m_{0}(t)$ resp. $\tilde{m}(t)$ for $g=1.25$
and $h=0.06$, $0.10$ and $0.12$, showing both the real time dependence
and its Fourier power spectrum (FPS). Frequencies are shown in units
of $J$.}
\end{figure*}
The three components of longitudinal magnetisation can be computed
as 
\[
m_{\alpha}(t)=\langle\Psi(t)|P_{i}^{\alpha}|\Psi(t)\rangle
\]
Due to translation invariance they are independent of the spatial
position $i$ and from the definitions (\ref{eq:Palpha_and_Ptilde})
and the residual symmetry $\mathbb{Z}_{2}$ they satisfy
\[
m_{1}(t)=m_{2}(t)=-\frac{m_{0}(t)}{2}\ .
\]
Transverse magnetisation can be defined as 
\[
\tilde{m}(t)=\langle\Psi(t)|\tilde{P}_{i}|\Psi(t)\rangle\ .
\]
An example of their time evolution is shown in Fig. \ref{fig:Time-evolution-of}.
Following \cite{Ising_confinement} and \cite{Ising_para}, the Fourier
spectra of their time series can be used as to determine the post-quench
quasi-particle spectrum via a sort of ``quench spectroscopy''. The
power spectra were obtained using FFT with an angular frequency resolution
$d\omega=2\pi/T\simeq.157$ and are also illustrated in Figs. \ref{fig:Time-evolution-of}.
They agree well with the predicted quasi-particle gaps; note that
due to the $\mathbb{Z}_{2}$ symmetry of the initial state preserved
by the post-quench Hamiltonian (\ref{eq:potts_chain_hamiltonian_broken-1}),
only $\mathcal{C}$-even states are visible. The second peak in the
power spectrum which appears above the critical value of $h$ is the
signature of a new bound state, in agreement with the predicted spectrum
from exact diagonalisation.

We remark that the post-quench state has a finite energy density,
which induces corrections in the quasi-particle spectrum and introduces
a finite life-time. The presence of well-defined quasi-particle peaks
close to the values extracted from the spectrum of the zero-density
system demonstrates that the post-quench dynamics can be described
in terms of the quasi-particle picture despite the non-integrability
of the system, similarly to the Ising case considered in \cite{Ising_para}. 

\subsection{Time evolution of entanglement entropy}

\begin{figure}
\begin{centering}
\includegraphics{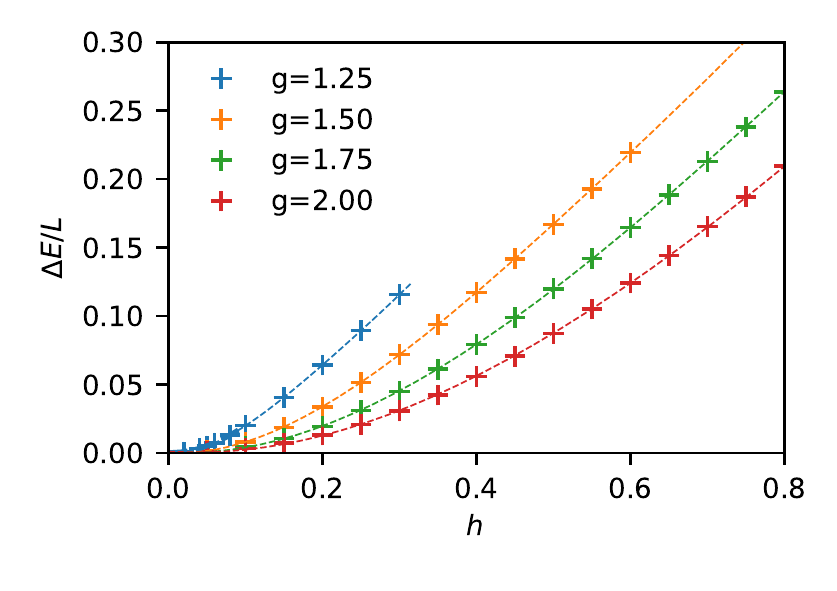}
\par\end{centering}
\caption{\label{fig:Density-of-energy} Density of energy (in units of $J/a$,
with $a$ denoting the lattice spacing) released by the quench as
a function of $h$. }
\end{figure}

Now we return to the effect observed in Fig. \ref{fig:The-mean-entanglement}.
Starting our discussion with the decreasing trend just before $h_{\text{min}}$,
we note that while presently a full quantitative understanding is
missing, the qualitative picture is clear. The initial increase in
$\overline{\partial_{t}S}$ comes from the energy density of the quench
increasing with $h$ as shown in Fig. \ref{fig:Density-of-energy}.
The subsequent decline in $\overline{\partial_{t}S}$ is consistent
with the quasi-particle gaps increasing, and the Lieb-Robinson velocities
decreasing with $h$ as demonstrated in Figs. \ref{fig:Quasi-particle-gaps-}
and \ref{fig:Lieb-Robinson-velocities-}, respectively. In fact for
very large values of $h$ the dynamics of the whole chain stops since
the Potts spins are locked in the direction of $h$, suppressing the
propagation of excitations along the chain as shown in the next Subsection.

Now let us consider the reversal of the decreasing trend, which happens
at the position $h_{\text{min}}$ of the local minimum in $\overline{\partial_{t}S}$.
The value of $h_{\text{min}}$ can be compared to the threshold $h_{\text{crit }}$
for the excited even quasi-particle $B$:
\begin{center}
\begin{tabular}{|c|c|c|c|c|}
\hline 
$g$ & $1.25$ & $1.5$ & $1.75$ & $2.0$\tabularnewline
\hline 
\hline 
$h_{\text{min}}$ & $0.10$ & $0.28$ & $0.49$ & $0.72$\tabularnewline
\hline 
$h_{\text{crit}}$ & $0.10$ & $0.28$ & $0.48$ & $0.71$\tabularnewline
\hline 
\end{tabular}
\par\end{center}

It is clear that the two positions coincide within numerical accuracy
(corresponding to the number of digits shown in the above table) for
small values of $g$, while $h_{\text{min}}$ is slightly larger than
$h_{\text{crit }}$ for larger $g$. This is the same pattern as observed
for the Ising spin chain in \cite{Ising_para}, and it can be explained
in the same way. 

Firstly, the appearance of the new quasi-particle species $B$ for
$h>h_{\text{crit }}$ leads to a steep increase in the entanglement
entropy production due to the contribution of Gibbs mixing entropy
arising from species information carried by the post-quench quasi-particles.
While a full quantitative description is lacking at the moment, there
is a simple argument using quasi-particle pair production rates determined
in the scaling Ising field theory that shows that the presence of
mixing entropy can lead to an order-of-magnitude increase in entropy
production. This argument is presented in \cite{Ising_para}, and
we do not repeat it here in detail.

Secondly, the fact that $h_{\text{min}}-h_{\text{crit}}$ is non-zero
and grows with $g$ can also be easily understood. Note that before
particle $B$ appears, $\overline{\partial_{t}S}$ has a decreasing
trend which is reversed by the appearance of $B$. However, the rate
of production of pairs containing $B$ is expected to rise only gradually.
The reason is when $B$ is only very weakly bound, the finite density
post-quench medium easily destabilizes it. So the higher the value
of $h_{\text{crit}}$, the larger is the quench when $B$ appears,
leading to a higher destabilizing effect of the post-quench medium
to be overcome. Since $h_{\text{crit}}$ increases with $g$, the
difference $h_{\text{min}}-h_{\text{crit}}$ is also expected to increase
with $g$ as well. This is indeed what was observed both here in the
Potts case, and also the Ising case considered in \cite{Ising_para}.

\subsection{Large $h$ behaviour}

\begin{figure}[t]

\begin{centering}
\includegraphics{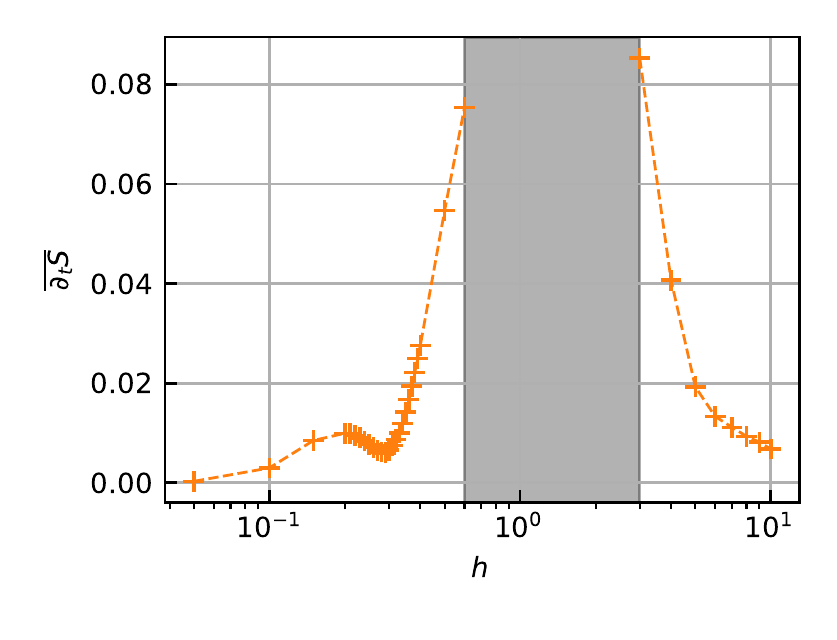}
\par\end{centering}
\caption{\label{fig:Entanglement-entropy-production} Entanglement entropy
production rate $\overline{\partial_{t}S}$ for $g=1.5$, including
the regime of large $h>0$. The shaded region corresponds to a parameter
range where entropy growth was so fast that $\overline{\partial_{t}S}$
could not be evaluated from iTEBD as it was impossible to follow the
dynamics for long enough times.}
\end{figure}

Recalling the Hamiltonian (\ref{eq:potts_chain_hamiltonian_broken-1})
we see that for large $h\gg g$ the spins of the chain are essentially
frozen in direction $0$. This is consistent with the increasing gaps
and decreasing velocities for the excitations shown in Figs. \ref{fig:Quasi-particle-gaps-}
and \ref{fig:Lieb-Robinson-velocities-}. For a very large value of
$h$, the dynamics slows down and $\overline{\partial_{t}S}$ goes
to zero, as shown in Fig. \ref{fig:Entanglement-entropy-production}. 

\subsection{The regime $h<0$ }

\begin{figure}[t]
\begin{centering}
\includegraphics[scale=0.93]{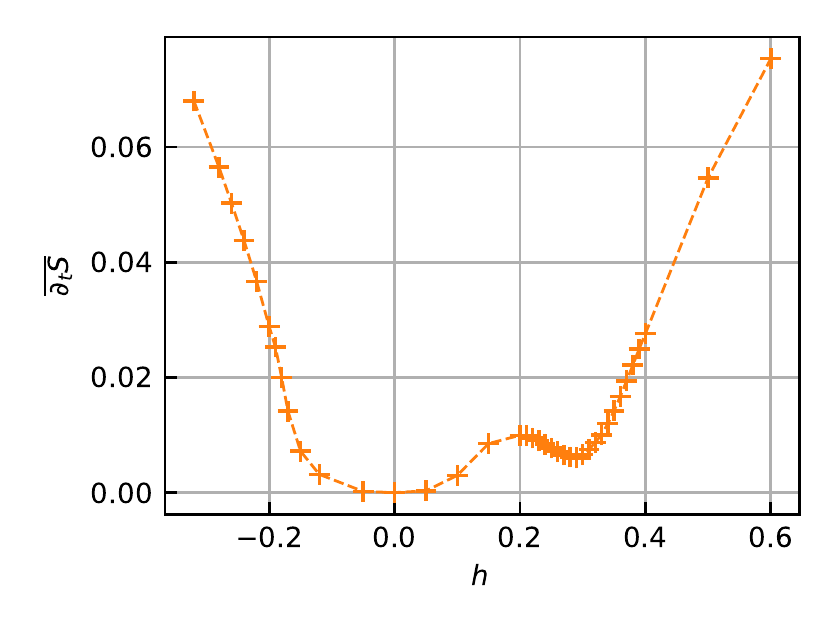}
\par\end{centering}
\caption{\label{fig:Entanglement-entropy-production-1} Entanglement entropy
production rate $\overline{\partial_{t}S}$ for $g=1.5$ including
the range $h<0$}
\end{figure}

\begin{figure}
\begin{centering}
\includegraphics{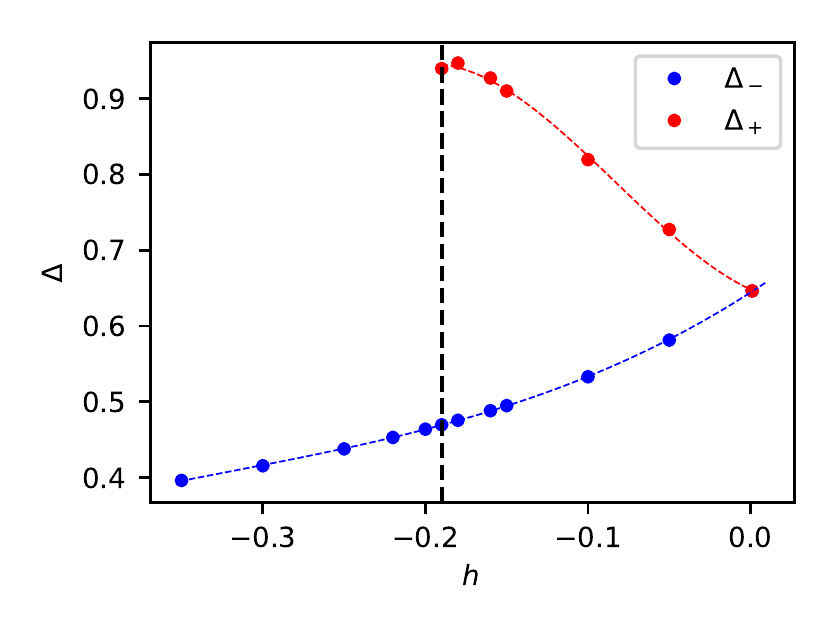}\includegraphics{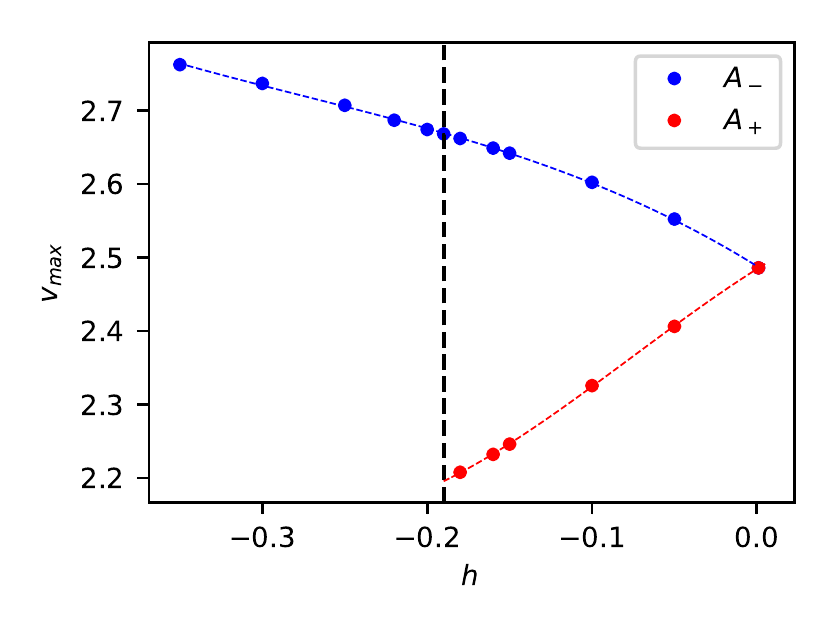}
\par\end{centering}
\caption{\label{fig:Quasi-particle-gaps-and} Quasi-particle gaps $\Delta_{\pm}$
(in units of $J$) and Lieb-Robinson velocities $v_{\mathrm{max}\pm}$
(in units of $Ja$, where $a$ is the lattice spacing) for $h<0$
(with $g=1.5$). The vertical dashed line shows the threshold value
$h_{-}\approx-0.19$ beyond which $A_{+}$ becomes unstable and decays
into two $A_{-}$ quasi-particles.}
\end{figure}

When $h<0$, no freezing of the dynamics occurs for $h\ll-g$. The
reason is that although a large negative $h$ freezes direction $0$,
the energetically favoured directions $1$ and $2$ remain degenerate
and so the chain effectively enters an Ising regime where $\overline{\partial_{t}S}$
grows monotonously with the amount of energy injected into the system
as shown in Fig. \ref{fig:Entanglement-entropy-production-1}. Examining
the quasi-particle threshold shows that here $A_{-}$ is lighter than
$A_{+}$ and there is even a threshold value $h_{-}$($\approx-0.19$
for $g=1.5$) below which $\Delta_{+}>2\Delta_{-}$. Therefore, for
$h<h_{-}$ the excitation $A_{+}$ becomes unstable and decays into
a pair of $A_{-}$ particles; as a result, the number of available
species decreases. 

Turning to the details of the quasi-particle spectrum, a direct calculation
using exact diagonalisation (as described in Section \ref{sec:Spectrum-of-the})
shows that the quasi-particle gap $\Delta_{-}$ rapidly decreases,
while the Lieb-Robinson velocities $v_{\mathrm{max}-}$ increases
when $h$ becomes more negative as shown in Fig. \ref{fig:Quasi-particle-gaps-and}.
Coupled with the rapid increase of the energy density injected in
the quench very similar to the $h>0$ domain (starting with a quadratic
rise and having a linear asymptotics for large $|h|$, cf. Fig. \ref{fig:Density-of-energy}),
this explains the rapid rise in $\overline{\partial_{t}S}$. Note
that even though $A_{+}$ becomes unstable at $h_{-}$, it barely
has any effect on entropy generation. The reason is that the gap and
Lieb-Robinson velocity of $A_{+}$ behave in an opposite way compared
to $A_{-}$, so with $h$ becoming more negative the share of $A_{+}$
in the entropy production decreases rapidly. By the point when $h$
passes through $h_{-}$, the only observable effect of $A_{+}$ becoming
unstable is a hint of an inflection point in the dependence of $\overline{\partial_{t}S}$
on $h$ at the threshold (cf. Fig. \ref{fig:Entanglement-entropy-production-1}).

\section{Conclusions and outlook \label{sec:Conclusions}}

In this work we considered quantum quenches in the paramagnetic phase
of the quantum Potts spin chain corresponding to switching on a longitudinal
magnetic field. We have demonstrated that the entanglement entropy
production rate shows a deep relation with the quasi-particle spectrum.
In particular, the mean entanglement entropy production rate $\overline{\partial_{t}S}$
is greatly enhanced by the appearance of a new quasi-particle species
in the spectrum, which is a manifestation of Gibbs mixing entropy
corresponding to species information. These findings are completely
consistent with the results obtained for the Ising case in \cite{Ising_para},
showing the general nature of the effect and confirming its interpretation
as a non-equilibrium manifestation of the so-called Gibbs paradox.
We have also shown that for very large $h>0$ the entropy production
rate decreases towards zero, which can be understood from the freezing
of spin chain dynamics.

In contrast with the Ising model, for the Potts case the domain $h<0$
has a different physics since the spin dynamics does not freeze for
any magnitude of $h$, and indeed in that domain we observed a monotonous
increase of $\overline{\partial_{t}S}$ with $|h|$, the qualitative
details of which again could be fully understood from the quasi-particle
spectrum.

As we noted, a detailed quantitative description of the entanglement
entropy production is not yet available. All the available evidence
shows that the quenches considered here are of sufficiently small
density to admit an essentially semi-classical quasi-particle description
following the picture proposed by Calabrese and Cardy in their seminal
works \cite{calabrese-cardy,calcard_entropy}. There has been substantial
recent work aiming at extending the quasi-particle description beyond
the simple picture of production of independent pairs, such as to
cases with no pair structure \cite{no_pair}, and initial states with
correlated pairs \cite{correlated_pairs}.

For a quantitative prediction of entanglement entropy production,
the main missing ingredient is sufficiently detailed knowledge of
the quasi-particle production rates as functions of the quench parameter
$h$. Presently such information is only available for the Ising case
and even there only in the scaling field theory limit \cite{e8_quenches}.
Once the amplitudes are available, one can try to develop a theory
for the entanglement entropy production following the lines of \cite{no_pair}.
Albeit in contrast to the case in \cite{no_pair} the system we consider
has interacting quasi-particles, it seems likely that in the regime
of transverse field $g$ close enough to the critical value $1$,
the post-quench density is small enough so that effects of interactions
between the quasi-particle do not affect substantially the post-quench
time evolution once the particles were created, and therefore one
could achieve at least a semi-quantitative description if the quasi-particle
production rates can be obtained by some means. 

\subsection*{Acknowledgments}

The authors are grateful to M. Kormos for comments on the manuscript.
This research was supported by the National Research Development and
Innovation Office (NKFIH) under a K-2016 grant no. 119204, and also
by the BME-Nanotechnology FIKP grant of EMMI (BME FIKP-NAT). G.T.
was also supported by the Quantum Technology National Excellence Program
(Project No. 2017-1.2.1-NKP-2017- 00001).

\end{document}